# Autonomous self-evolving research on biomedical data: the DREAM paradigm

Luojia Deng[a,b*], Yijie Wu[a,b*], Yongyong Ren[b*], Hui Lu[a,b#]

[a] Department of Bioinformatics and Biostatistics, School of Life Sciences and Biotechnology, Shanghai Jiao Tong University, Shanghai, China

[b] SJTU-Yale Joint Center for Biostatistics and Data Science, Technical Center for Digital Medicine, National Center for Translational Medicine, Shanghai Jiao Tong University, Shanghai, China

[*] These authors contributed equally to this work.

[#] Corresponding authors: Dr. Hui Lu, 800 Dongchuan Road, Minhang District, Shanghai, China (email: huilu@sjtu.edu.cn)

## Abstract

In contemporary biomedical research, the efficiency of data-driven approaches is hindered by large data volumes, tool selection complexity, and human resource limitations, necessitating the development of fully autonomous research systems to meet complex analytical needs. Such a system should include the ability to autonomously generate research questions, write analytical code, configure the computational environment, judge and interpret the results, and iteratively generate more in-depth questions or solutions, all without human intervention. In this study, we developed DREAM, the first biomedical Data-dRiven self-Evolving Autonomous systeM, which can independently conduct scientific research without any human involvement. Utilizing a clinical dataset and two omics datasets, DREAM demonstrated its ability to raise and deepen scientific questions, with difficulty scores for clinical data questions surpassing top published articles by 5.7% and outperforming GPT-4 and bioinformatics graduate students by 58.6% and 56.0%, respectively. Overall, DREAM has a success rate of 80% in autonomous clinical data mining. Certainly, human can participate in different steps of DREAM to achieve more personalized goals. After evolution, 10% of the questions exceeded the average scores of top published article questions on originality and complexity. In the autonomous environment configuration of the eight bioinformatics workflows, DREAM exhibited an 88% success rate, whereas GPT-4 failed to configure any workflows. Using the clinical dataset as an example, DREAM was over 10,000 times more efficient than the average scientist with a single computer core, and capable of revealing new discoveries. As a self-evolving autonomous research system, DREAM provides an efficient and reliable solution for future biomedical research. In the future, this research paradigm may also have a revolutionary impact on other data-driven scientific research fields.

## Introduction

The biomedical scientific research paradigms are typically categorized into hypothesis-driven and data-driven approaches. Hypothesis-driven research begins with a specific question and seeks to utilize data to address it. On the other hand, data-driven research derives questions from vast amounts of existing data and discovers valuable insights through data mining. Currently, the volume of data in the biomedical research field is exceedingly large[1], making it extremely challenging to raise numerous valuable scientific questions in a short time, and the variety of analytical tools available also leads to complexity and confusion in their selection and application[2]. The efficiency of data mining is often constrained by researchers' varying proficiency in programming, experience with tools, and understanding of parameter settings, leading to significant disparities in performance. Therefore, enhancing the autonomy level of data mining is crucial for improving overall research efficiency.

In recent years, the rapid development of large language models (LLMs)[3-5], particularly the release of ChatGPT 3.5, has garnered widespread attention. The launch of GPT-4 by OpenAI[6], has spurred various LLMs-based research systems. In the field of chemistry, systems such as Coscientist[7] and ChemCrow[8] utilize LLMs to aid scientific investigations. In biology, systems like BIA[9] and Bio-copilot[10] have demonstrated its abilities in assisting bioinformatic analysis as 'co-pilot'. In clinical research, ChatGPT ADA[11] has proven effective in helping users develop machine learning models. Similarly, data-to-paper[12] can automatically propose some simple hypothesis testing questions from data and complete the analysis. Furthermore, systems like DS-Agent[13] and MLAgentBench[14] can help researchers build models and perform data analysis tasks. These systems provide new avenues for scientific research, enhancing research efficiency to a certain extent.

Despite advancements in LLMs-based research agent systems, a fully autonomous data-driven research system has not yet been realized. Such a system should include the ability to autonomously generate research questions, write analytical code, configure the computational environment, judge and interpret the results, and iteratively reflect and generate more in-depth research questions or solutions, all without human involvement. However, existing systems still rely heavily on human interventions during the operational processes and cannot eliminate human involvement. Specifically, Coscientist[7] cannot autonomously generate questions nor configure the computational environment. ChemCrow[8] requires predefined tasks, manual environment setup, and lacks iterative reflection. BIA[9] requires human-defined questions and environment configuration, without advanced iterative functionality. Bio-Copilot[10] necessitates human interaction at every step, including question generation, environment setup, and result evaluation. Systems like DS-Agent[13] and MLAgentBench[14] cannot autonomously generate questions, configure the environment, or perform advanced iterative reflection. Data-to-paper[12] also lacks in autonomous environment configuration, result evaluation, and iterative capabilities. Additionally, all these systems are limited to Python, without extension to other languages like Shell or R. Thus, developing a truly autonomous research system that minimizes reliance on human involvement is crucial for large scale research and achieving significant breakthroughs in scientific progress.

In this study, we propose a completely autonomous data-driven self-evolving research system named DREAM. The system operates entirely without any form of human intervention, capable of functioning 24/7 and scaling up its core components as needed to enhance efficiency. Certainly, human involvement can be integrated at any stage of the process to achieve more personalized goals, while offering more comprehensive functions than the current 'co-pilot' systems. DREAM can autonomously interpret data, generate scientific questions, acquire necessary variables, plan tasks, write code, configure experimental environments, evaluate results, correct erroneous code, interpret and validate outcomes, and propose more in-depth questions that need to be further explored based on the analysis results. In the following section, we first provide a detailed overview of generation and evolution of scientific questions based on biomedical data, then present two novel core modules of DREAM: (1) configuration of experimental environments; (2) evaluation of the analysis workflow and results. To validate the effectiveness and interaction of each module, we conducted ablation experiments and analyzed the roles of prompts through 'basic prompt' experiments. Furthermore, we demonstrate the potential applications of DREAM in data mining, particularly in uncovering new scientific discoveries. The results indicate that DREAM can significantly enhance research efficiency and autonomously generate and evolve innovative scientific questions, and verify them by computing, showcasing its substantial potential in life sciences and medical research.

# Results

## Architecture of DREAM

Within the data-driven scientific research paradigm (Figure 1a), where there are four core steps: 'QUestion', 'codE', 'coNfIgure' and 'jUdge' (UNIQUE), we propose the DREAM leveraging LLMs, capable of fully autonomous biomedical research without any form of human involvement. As illustrated in Figure 1b, the architecture of DREAM encompasses all aspects of the UNIQUE paradigm, comprising eight steps and ten major modules. With structured biomedical data, taking omics and clinical data as examples, DREAM autonomously interpret information (dataInterpreter) from data, raise research questions (questionRaiser), handle tasks such as screening relevant variables (variableGetter), planning analysis tasks and steps (taskPlanner), writing analysis code (codeMaker), configuring the computing environment (dockerMaker), running and debugging the code (codeDebugger), and judging (resultJudger) and interpreting results based on the data and research questions (resultAnalyzer). After analyzing a scientific question, the system enters a cycle of self-reflection, iteration, and evolution, raising deeper questions (deepQuestioner) based on the analysis results, and repeating this process to continuously advance scientific research. An additional module is also available for seeking existing evidence to validate the obtained results (resultValidator) as shown in Figure 1c. Our demonstration website is available at https://bmap.sjtu.edu.cn/dream.

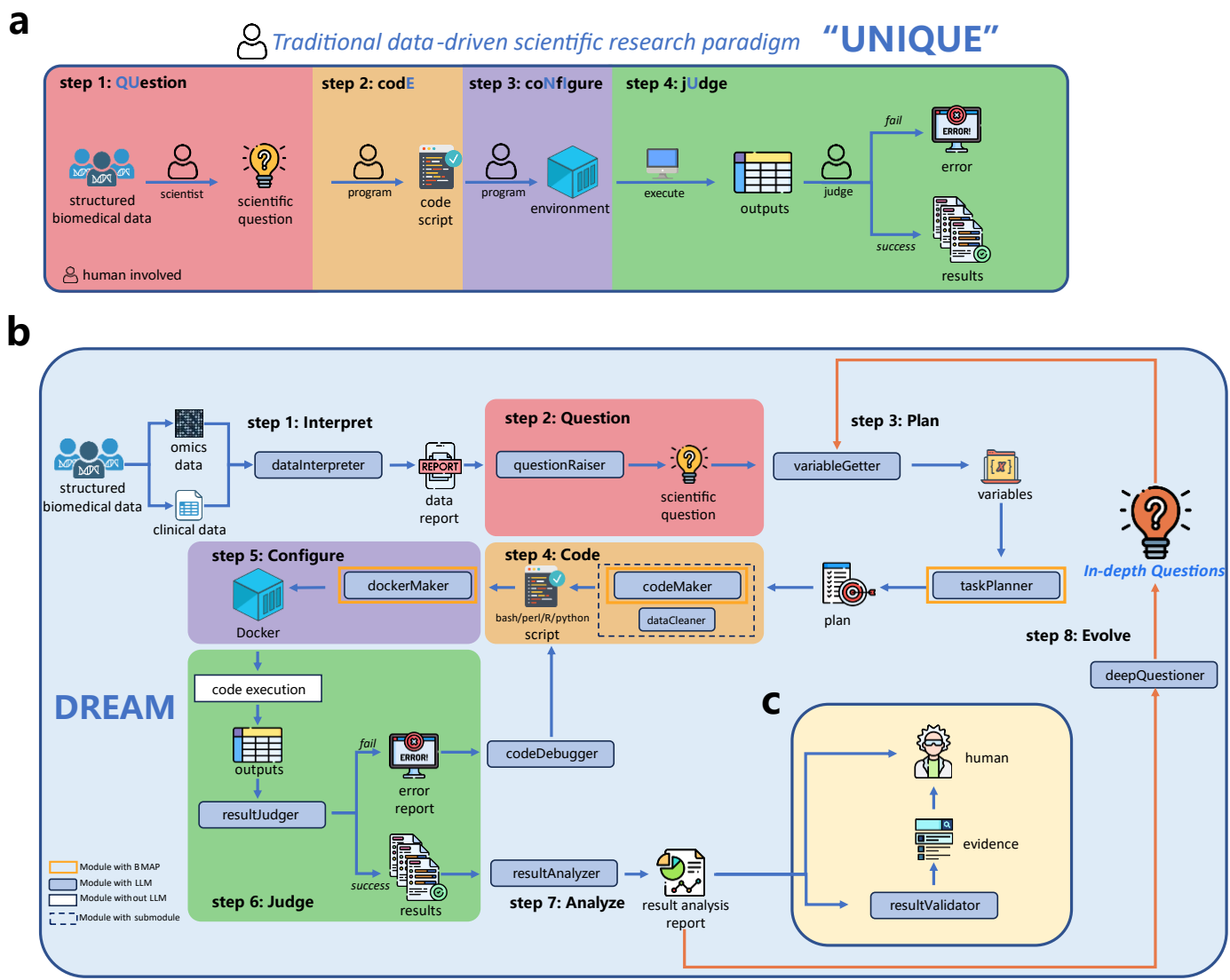

**Figure 1. The system's architecture. a.** Traditional data-driven scientific research paradigm (UNIQUE). From raising research question to coding, environment configuring, and result judgement and evaluation, each step requires human involvement. **b.** DREAM based on LLMs. The architecture of DREAM comprises multiple modules with different functions. Orange box outlines represent that the module utilizes the external tool library BMAP[15]. **c.** External validation of the questions and analysis result provided by DREAM.

## Self-evolution of the system

A notable gap exists among current LLMs-based research systems, in autonomously generating research questions[7-10,13]. These systems still rely on human designation of tasks. In contrast, the questionRaiser module (Figure 2a) within DREAM is capable of independently raise scientific questions based on structured biomedical data. Moreover, no existing system can iteratively generate new questions, whereas the deepQuestioner module in DREAM can propose more in-depth questions by understanding previous questions and results, enabling DREAM to continuously learn and evolve. To assess the quality of the generated questions, we evaluated the difficulty and quality scores[16] from different aspects. Detailed evaluation criteria are in the supplementary file. To evaluate the evolutionary capability of DREAM, we conducted four rounds of evolution on clinical data. To compare with top experts, we selected questions from the top articles in the clinical dataset we used and applied the same scoring criteria.

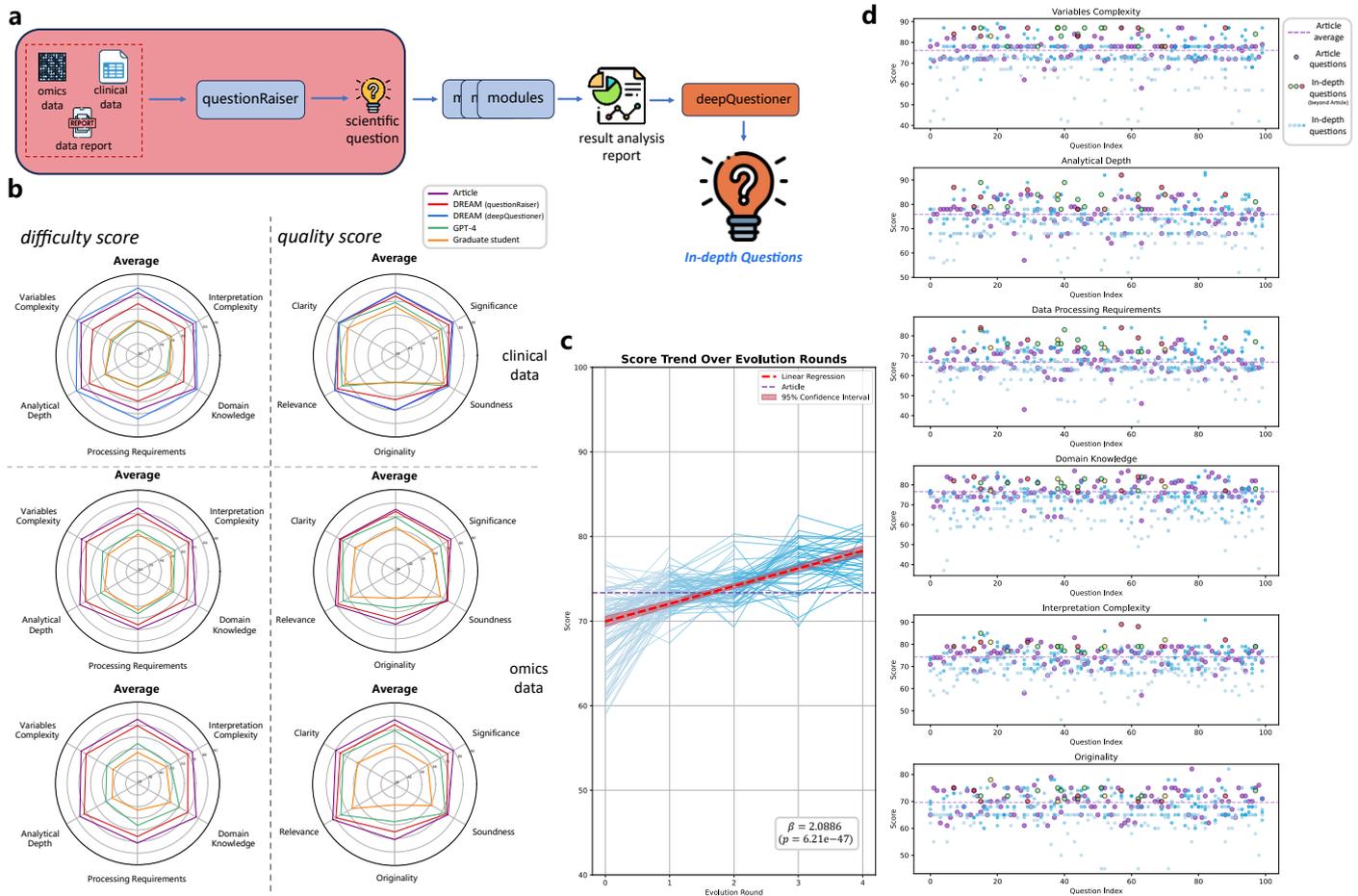

**Figure 2. DREAM's self-evolution on raising scientific questions. a.** The process of raising and deepening scientific research questions. **b.** Comparison of difficulty scores and quality scores across dimensions for core scientific questions summarized from published articles and scientific questions raised by DREAM (questionRaiser), DREAM (deepQuestioner), GPT-4, and bioinformatics graduate students. **c.** The trend of question score changes in DREAM during self-evolution, with the purple dashed line in the graph representing the average score of the questions in the article. The red dashed line and deep red area represent the regression line and 95% confidence interval, respectively. **d.** Comparison of key dimensions after question self-evolution loop. The purple dots represent the scores of the questions in the article. The four blue gradient dots represent the question scores in the four rounds of DREAM self-evolution. The yellow, green, and red dots indicate questions where the scores of six key indicators in the second, third, and fourth rounds of evolution exceed the average level of the article questions.

Regarding difficulty scoring (Figure 2b), in the initial round, core scientific questions from published top-tier articles scored the highest across nearly all dimensions and datasets, reflecting the significant complexity of leading expert-raised and peer reviewed questions. Although these questions did not achieve perfect scores, this could be attributed to the conservative scoring by LLMs. DREAM (questionRaiser) posed questions with difficulty second only to published articles, indicating a relatively high level of difficulty. Questions generated by GPT-4 and graduate students were similar in difficulty, with different strengths across various datasets and dimensions, but their overall scores were significantly lower than those of published articles and DREAM. Furthermore, in clinical data, through self-evolution, the average difficulty score of questions posed by DREAM (deepQuestioner) exceeded those by GPT-4 and bioinformatics graduate students by 58.6% and 56.0%, respectively, and surpassed those in published articles by 5.7%. This highlights the advantage of self-reflection and iteration, showing that after four rounds of evolution, the system can surpass leading experts.

Regarding quality scoring (Figure 2b), although DREAM (questionRaiser) scored lower than published articles, the difference was smaller than that in difficulty, with no significant difference in clarity in clinical data and soundness in omics data. Additionally, in clinical data, the questions generated by DREAM (deepQuestioner) scored higher than the questions from published articles, though not significant. DREAM (deepQuestioner) showed a 12.3% improvement in originality compared to DREAM (questionRaiser), and exceeded GPT-4 and bioinformatics graduate students by 41.4% and 40.6%, respectively. GPT-4's lower score might be attributed to the simplicity of its questions, leading to lower originality scores, but high core in clarity and soundness. Graduate students had the lowest overall quality score, indicating lower quality of questions formulated in a short time. In addition, the dataInterpreter in DREAM greatly increases the difficulty of the subsequent questions raised by QuestionRaiser. Without dataInterpreter, the difficulty score of questions in clinical data is significantly lower in all dimensions than the questions proposed by graduate students. The trend in omics data was similar to clinical data, but DREAM's scores were closer to published questions, and graduate student performed markedly poorer, highlighting the difficulty for human in posing reasonable questions in more complex omics data.

To explore the capability and effectiveness of system evolution, we performed linear regression on the scores from four rounds of evolution, with significant regression coefficient (Figure 2c, P-value<0.05). After two rounds of evolution, DREAM's questions surpassed the level of those from top-tier articles, showing an overall upward trend with some fluctuations. Key dimensions (Figure 2d) indicate that in six key dimensions, most questions exceeded the average level of published articles after four rounds of evolution. Furthermore, 10% of the questions surpassed the level of published articles in all key dimensions, and 17 out of 25 (68%) high-level questions were successfully addressed. This demonstrates DREAM's potential to surpass top human scientists through iterative self-reflection and evolution.

Overall, the scientific questions in top-tier published articles are well-balanced in difficulty and quality, with high scores across all dimensions. DREAM also performed well, with deepQuestioner generating in-depth questions that surpassed the difficulty scores of questions in published articles and matched their quality scores. Questions generated by GPT-4 and graduate students performed well in certain quality dimensions but had lower overall difficulty. These findings provide crucial insights into the application and evaluation of scientific questions from different sources in research.

## Core modules functions and performance

### Computation environment configuration

In existing agent research systems, pre-installed environments not only limit the scope of software usage, but also require manual configuration, thereby limiting the breadth and depth of research. Unlike other manually pre-configured environmental systems, the dockerMaker module in DREAM provides the ability to autonomously configure the runtime environment required for analyzing workflows.

To evaluate the functionality and feasibility of the dockerMaker module, we conducted experiments on the environment configuration of eight common analysis workflows in the field of biomedicine, particularly in bioinformatics (Table 1). The scripts were written in multiple programming languages, such as shell, Perl, R, and Python. The evaluation metrics included whether the workflow installation was successful and the proportion of successfully installed software. Specifically, we compared the effectiveness of DREAM with manual installation and GPT-4 installation with a basic framework.

From Table 1, it can be observed that in terms of workflow installation, DREAM achieved the workflow success rate of 88%, and the success rate of software installation is 99%. In contrast, the senior human installer failed three workflows, resulting in an installation success rate of 63%, while the junior human installer failed five workflows, with a success rate of only 38%. The GPT-4 with basic framework did not successfully complete the installation of any workflow, yielding a success rate of 0%. To sum up, dockerMaker in DREAM demonstrated its excellent performance in computational environment configuration, showing the capability to accommodate a diverse range of complex analysis workflows. These results underscore the critical role of the dockerMaker module in autonomous scientific research, particularly in enhancing workflow setup efficiency and reliability.

Table 1. Comparison of analysis workflow environment configurations

| Installer | Workflow success rate | Software success rate |
|---|---|---|
| DREAM | 88% | 99% |
| Human (senior) | 63% | 93% |
| Human (junior) | 38% | 81% |
| GPT-4 | 0% | 52% |

**Judgement of analysis success**

Currently, most research agent systems lack autonomous result-judging capabilities and typically rely on human experts[9,10,12,14]. However, the core module resultJudger in DREAM replaces the labor-intensive Judge step in the UNIQUE paradigm and achieves automated judgements on analysis. We generated and attempted to solve 100 questions based on clinical data, inviting human experts to evaluate whether these questions were answered. This served as a performance evaluation for resultJudger, which was then compared with GPT-4.

As shown in Figure 3a and 3b, DREAM (resultJudger) achieved scores above 0.9 in precision, recall, specificity, F1-score and AUC, all of which are higher than those of GPT-4, while GPT-4 only performed adequately in recall. Furthermore, in scientific research, researchers are usually more concerned with a low false positive rate, while a slightly higher tolerance for false negatives is acceptable. That is, it is more critical that true errors are not judged as correct, even if some true positives are judged as incorrect. The high precision (0.926 vs 0.750) and high specificity (0.917 vs 0.667) in the results demonstrate that DREAM (resultJudger) exhibits human-like judgement capabilities.

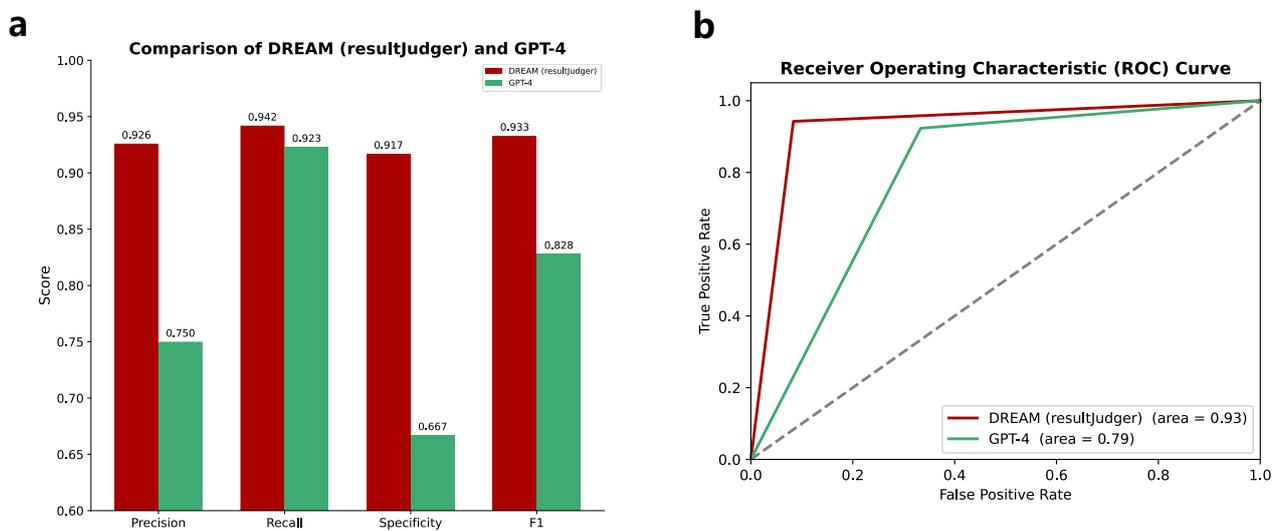

**Figure 3.** DREAM' capabilities in judging whether a question has been successfully analyzed and answered. **a.** Comparison of precision, recall, specificity and F1-score between DREAM (resultJudger) and GPT-4. **b.** Receiver Operating Characteristic (ROC) Curve of DREAM (resultJudger) and GPT-4.

In summary, although DREAM (resultJudger) may occasionally misjudge unanswered questions as resolved, it represents a notable advancement in autonomous scientific research by improving judgment accuracy and efficiency. By reducing reliance on human experts, DREAM (resultJudger) optimizes the research process, ensures greater consistency and reproducibility, and provides a powerful tool for more advanced scientific autonomy in the future.

**Ablation and basic prompt study**

To investigate the roles of modules within DREAM and the effectiveness of their prompts, we conducted ablation experiments and a series of 'basic prompt' experiments. For 100 clinical questions posed by questionRaiser, Figure 4 illustrates the pass rates of the original code (Original) and after four debugging cycles (Debug 1 to Debug 4). All evaluations were conducted by resultJudger in DREAM.

In the ablation study, we assessed the DREAM's performance with the taskPlanner, variableGetter, and dataCleaner modules individually removed. Additionally, we evaluated DREAM (GPT-3.5) and a baseline system under the traditional UNIQUE framework, where GPT-4 replaced human roles. Only DREAM achieved a pass rate exceeding 50% in the initial round and consistently outperformed other groups across all debugging rounds (Figure 4a). The UNIQUE (baseline) system performed the worst, with a significantly lower pass rate. Notably, DREAM (GPT-3.5) outperformed DREAM without the dataCleaner module in the initial round, highlighting the dataCleaner module's contribution to the original pass rate. Although the removal of variableGetter, taskPlanner, and dataCleaner modules significantly impacted the initial pass rates compared to the complete DREAM system, the impact diminished after several debugging rounds, with performance nearing that of the complete system.

The 'basic prompt' study involved replacing certain module prompts with basic prompts, the most basic task descriptions, while keeping other module prompts unchanged. 'Basic DREAM' refers to the system where all module prompts within DREAM were replaced with basic prompts. The results indicate that substituting any module's prompt with a basic prompt reduces the system's performance to varying degrees, and the fully basic DREAM system performs significantly worse than the other systems (Figure 4b). This suggests that while LLMs' powerful language capabilities allow DREAM to handle some questions with simple task descriptions, basic prompts are less effective for complex tasks compared to our meticulously designed prompts.

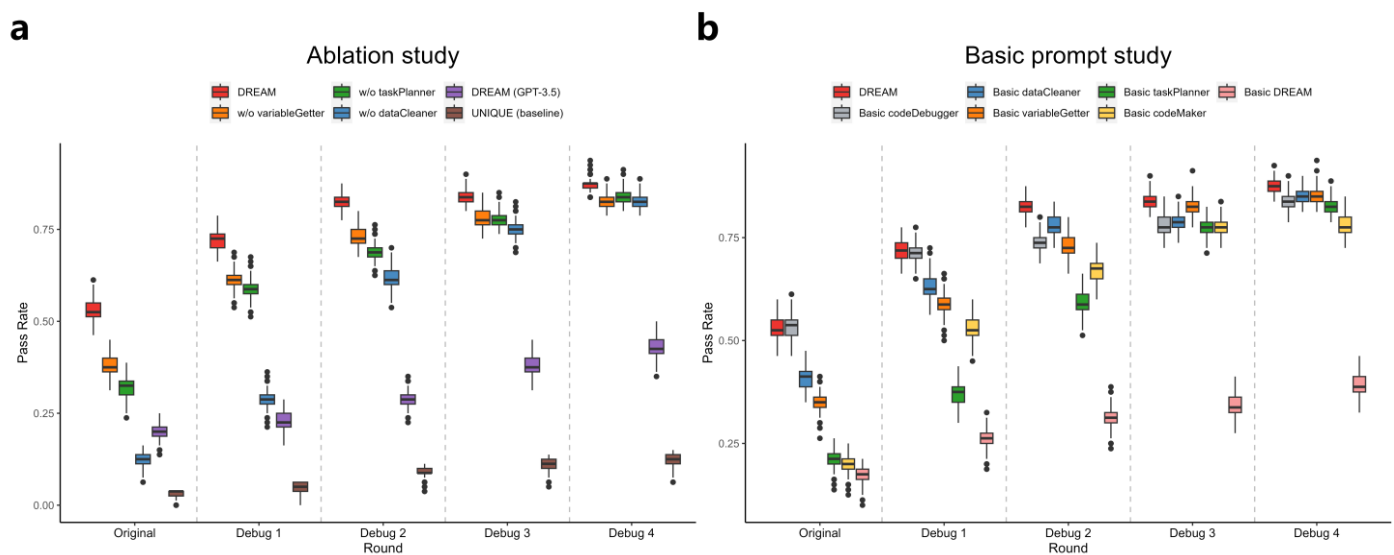

**Figure 4. Performance evaluation of DREAM modules and prompt configurations. a.** Comparison of question pass rates in different rounds with three ablatable modules removed. **b.** Comparison of question pass rates for different basic modules using basic prompts in different rounds.

These findings clearly emphasize the importance of each module and carefully crafted prompts in DREAM, providing valuable support for further optimization of autonomous scientific research systems.

## Case studies and new discoveries

### Pathogenic gene mutations identification

Genetic diseases, caused by changes in genetic material, fall into two major categories: germline and somatic mutations[17]. DNA sequencing technologies, particularly NGS-based WES and WGS, are widely used to detect various genetic diseases, including chromosomal abnormalities, copy number variations, and monogenic disorders[18-20].

In this case, WES data was provided to DREAM. Taking the question "What are the potential pathogenic mutations present in the exome data from the patient with genetic disease?" as an example, since this data lacks additional metadata, the variableGetter module was not utilized, and the process proceeded directly to the taskPlanner module. The planned steps include: 1) quality check for paired-end sequencing data; 2) trim adaptor sequences; 3) align reads to the reference genome; 4) convert SAM to BAM; 5) variant calling; 6) annotate variants and predict pathogenicity. The codeMaker module wrote the analysis script based on the steps above; dockerMaker configured the computational environment. After one round for debugging, DREAM successfully completed the analysis workflow and obtained the final ranked list of pathogenic gene mutations. The top-ranked gene mutation was *PRRT2* p.Arg217fs/c.649dupC, which matched the actual gene mutation, confirming the reliability of DREAM for bioinformatics analysis.

### New discoveries in clinical indicators

To validate the questions proposed by DREAM and the corresponding results, we created an automated resultValidator, specifically for clinical dataset we used. The resultValidator achieves a precision of 1.0, a recall of 0.95, and an F1-score of 0.974, indicating its ability to accurately determine whether a specific research question has already been studied in the dataset.

For the 100 clinical research questions raised by DREAM, the resultValidator found that 55% were determined to have not been researched yet. For instance, DREAM concluded that women with a college degree or higher have a 21% lower risk of experiencing a stroke compared to those with lower educational attainment. Existing studies have not explored this question in this dataset. However, similar conclusions have been found in other datasets. For example, in a prospective cohort study, Jackson et al. found that women with lower educational attainment had a 21% to 41% higher risk of stroke compared to those with a college degree or higher[21]. Furthermore, some studies have indicated that, compared to high-income countries, higher educational levels may not provide a protective effect against cardiovascular events in low- and middle-income countries, particularly among women[22]. This suggests that higher education may positively impact women's health by improving health behaviors and increasing access to healthcare.

**Enhancement of Research Efficiency**

Table 2. Comparison of Research Efficiency

| Researcher | Number of solved sub-questions per person-day mean (sd) |
|---|---|
| DREAM | 349.39 (252.74) |
| Top human scientists | 0.746 (0.483) |
| Human scientists | 0.032 (0.020) |

To evaluate the efficiency of human researchers in solving scientific questions, we collected all publications based on the clinical data. It was calculated that each author could solve approximately 0.032 sub-questions per day. For top scientists, the estimated number of sub-questions solved per person-day is around 0.746.

In comparison, DREAM demonstrates significantly higher research efficiency even in a single-core environment. According to experimental data, DREAM successfully solved around 1397.56 sub-questions in 24 hours. Our research team comprises four researchers, and for comparative purposes with human scientists, we divided this number by 4, resulting in 349.39 sub-questions per person-day. As illustrated in Table 2, the research efficiency of DREAM on a single core is approximately 10,000 times the average level of researchers and 468 times that of top-tier scientists. When more cores are utilized, this efficiency would increase linearly.

The significant boost in efficiency is primarily attributed to the autonomy and high computational capability of DREAM system in handling scientific questions. Unlike traditional research, which involves substantial time investment in thinking, calculating, and report writing, DREAM drastically reduces the time required for these tasks, thereby significantly enhancing research efficiency. This demonstrates the tremendous potential of autonomous scientific research systems in managing large-scale scientific data.

## Discussion

This work presents a biomedical data-driven self-evolving autonomous research system based on LLMs, named DREAM, showcasing its autonomous capabilities and efficiency in scientific research. DREAM can autonomously raise, answer and evolve scientific questions through the entire system process, significantly enhancing research efficiency and progress. Several case studies validate the system's effectiveness. Additionally, DREAM can discover new insights, such as the finding that women with a university degree or higher have a reduced risk of stroke, thereby providing new perspectives to existing literature and research. DREAM supports human involvement at any stage of the process for tailoring solutions to more personalized objectives while providing more comprehensive functions than the current 'co-pilot' systems. For example, if researchers already have a defined task, DREAM can assist in automatically completing the necessary analyses. For those who are technically proficient but less adept at generating research ideas, DREAM can also aid in brainstorming.

However, DREAM also has limitations. DREAM currently only supports structured data, and its ability to handle unstructured data such as images and videos needs improvement. Although dockerMaker is highly accurate in environment configuration, it does not significantly outperform senior human researchers in speed, highlighting the need for faster algorithms. Furthermore, the resultJudger has not yet achieved the expertise level of human experts in determining whether questions have been answered, which affects the overall performance and reliability of the system's results.

Overall, DREAM, as the first completely autonomous self-evolving research system based on data, demonstrates considerable potential in enhancing research efficiency, progress, and innovation. With advancements in computational power and algorithm optimization, DREAM will play a crucial role across a broader range of research fields, driving scientific research toward a more efficient and intelligent era. Future efforts should focus on improving the effectiveness of each module, enhancing the ability to process multimodal data, and optimizing self-reflection, iteration, and evolution mechanisms. These enhancements will expand the performance and application scope of DREAM, leading to significant breakthroughs and progress in scientific research.

# Methods

## A. Technical details of DREAM

**Self-reflection, self-iteration, and self-evolution**

Our system demonstrates strong capabilities for self-reflection, iteration, and evolution, primarily achieved through the cyclical operation of five modules: codeMaker, dockerMaker, resultJudger, codeDebugger, and deepQuestioner. When analyzing a scientific question, the codeMaker module generates the analysis code. After execution, the resultJudger evaluates the code. If the results are correct, the cycle ends. If the results are incorrect, the resultJudger provides feedback. Based on this feedback, the codeDebugger modifies the code and re-evaluates the revised version. In some cases, the issue may stem from the environmental configuration rather than the code itself. In such instances, the dockerMaker module becomes crucial. The dockerMaker reconfigures the runtime environment according to the resultJudger's feedback. It creates or adjusts the Docker environment to ensure the code runs in the appropriate settings. Once the environment is reconfigured, the code is executed again and evaluated by the resultJudger. Following the analysis, the deepQuestioner formulates further exploratory questions based on the initial question and the corresponding analysis results, thereby promoting ongoing data exploration. Through the collaboration of these modules, the system continuously self-reflects, iterates, and evolves.

## B. Evaluation metrics

**Question scoring**

To evaluate the scientific value of questions proposed by questionRaiser in comparison to scientific questions found in published articles, questions generated by GPT-4, and questions posed by graduate students in bioinformatics, we have developed a series of question scoring criteria based on two dimensions: difficulty and quality. The quality score primarily assesses whether a scientific question meets certain quality criteria, including clarity, feasibility, and originality. This standard is adapted from the scoring criteria provided by ResearchAgent[16], specifically optimized for biomedical questions, and serves as a foundational level of evaluation. The difficulty score, on the other hand, measures the complexity involved in solving a scientific question. This evaluation standard, newly designed by us, considers factors such as the required processing needed and the breadth of domain knowledge involved. This score represents a higher-level assessment of the challenge posed by the question.

**Statistical metrics**

To evaluate the performance of resultJudger and GPT-4 in assessing question responses and the performance of resultValidator in question validation, we employed the precision, recall, specificity, F1-score and AUC metrics, with higher scores indicating better performance.

## Acknowledgements

We acknowledge support from collaborators for providing data to test our DREAM program. We thank C. Fan, Y. Li, J. Sun, H. Zhao, X. Zhang for helpful discussions. We thank to Biorender and Flaticon, many of our icons are created with BioRender.com and Flaticon.com.

## Author contributions

L.D. performed the omics data analysis, statistical analysis, autonomous model design, and drafted the manuscript. Y.W. participated in the study design, clinical data analysis, visualization, and manuscript writing. Y.R. contributed to study design, methodology, data analysis, and manuscript revision. H.L. contributed to concept, methodology, manuscript revision, and supervised all aspects of the study. All authors reviewed and approved the final manuscript.

## Competing interests

The authors declare no competing interests.